\PassOptionsToPackage{sort&compress}{natbib}
\documentclass[unnumsec,webpdf,contemporary,large]{oup-authoring-template}%

\usepackage[caption=false]{subfig}
\usepackage{cleveref}
\graphicspath{{images/}{oup-authoring-template/images/}{figures/}{Fig/}}
\newcommand{\zenodoDOI}{23090828} % Zenodo version DOI of v0.1.0 (concept DOI 10.5281/zenodo.23090827)

\theoremstyle{thmstyleone}%
\theoremstyle{thmstyletwo}%
\theoremstyle{thmstylethree}%

\makeatletter
\def\ps@arxivfirst{%
  \let\@mkboth\@gobbletwo
  \def\@oddhead{{\sffamily\fontsize{8bp}{10bp}\selectfont
    Preprint.}\hfill}%
  \let\@evenhead\@oddhead
  \def\@oddfoot{}\let\@evenfoot\@oddfoot}
\def\ps@arxivrun{%
  \let\@mkboth\@gobbletwo
  \def\@evenhead{{\sffamily\fontsize{9bp}{10bp}\selectfont
    \thepage\ \raisebox{-1.5pt}{\rule{.3pt}{8pt}}\ napari-TurboBox (preprint)}\hfill}%
  \def\@oddhead{\hfill{\sffamily\fontsize{9bp}{10bp}\selectfont
    napari-TurboBox (preprint)\ \raisebox{-1.5pt}{\rule{.3pt}{8pt}}\ \thepage}}%
  \def\@oddfoot{}\let\@evenfoot\@oddfoot}
\makeatother

\begin{document}
\hypersetup{pdftitle={napari-TurboBox: Interactive Multi-Viewer Bounding Box Annotation for AI-Guided Volumetric Bioimaging},
  pdfauthor={Nils Friederich et al.}}
\journaltitle{Bioinformatics}
\DOI{DOI added during production}
\copyrightyear{2026}
\pubyear{2026}
\vol{XX}
\issue{x}
\access{Published: Date added during production}
\appnotes{Application Note}

\firstpage{1}

\title[napari-TurboBox]{napari-TurboBox: Interactive Multi-Viewer Bounding Box Annotation for AI-Guided Volumetric Bioimaging}

\author[1,2,$\ast$]{Nils Friederich\ORCID{0009-0003-9377-3140}}
\author[1,2]{Philip Kohle}
\author[1]{Maximilian Beichter\ORCID{0009-0009-9713-3590}}
\author[2]{Mona Wellh\"ausser\ORCID{0009-0002-4004-6508}}
\author[1]{Marcel Stahl}
\author[3]{Pascal Zwick\ORCID{0009-0009-6145-865X}}
\author[1]{Angelo Jovin Yamachui Sitcheu\ORCID{0009-0002-9384-6496}}
\author[1]{Thomas Lautenschlager\ORCID{0009-0002-3214-9363}}
\author[1]{David Exler\ORCID{0009-0005-4116-5287}}
\author[1]{Linus Schilpp}
\author[1]{Stephan Allgeier\ORCID{0000-0002-1218-448X}}
\author[4,5,$\oplus$]{Andreas Specker\ORCID{0009-0001-2391-3326}}
\author[2,6,$\oplus$]{Lennart Hilbert\ORCID{0000-0003-4478-5607}}
\author[1,$\oplus$]{Ralf Mikut\ORCID{0000-0001-9100-5496}}

\address[1]{\orgdiv{Institute for Automation and Applied Informatics (IAI)},
  \orgname{Karlsruhe Institute of Technology},
  \orgaddress{
  %\street{Hermann-von-Helmholtz-Platz 1}, 
  %\postcode{76344},
  %\state{Baden-W{\"u}rttemberg},
  \country{Germany}}}

\address[2]{\orgdiv{Institute of Biological and Chemical Systems (IBCS)},
  \orgname{Karlsruhe Institute of Technology},
  \orgaddress{
  %\street{Hermann-von-Helmholtz-Platz 1}, 
  %\postcode{76344},
  %\state{Baden-W{\"u}rttemberg},
  \country{Germany}}}

\address[3]{\orgdiv{Intelligent Systems and Production Engineering},
  \orgname{FZI Research Center for Information Technology},
  \orgaddress{
  %\street{Haid-und-Neu-Stra{\ss}e 10--14}, 
  %\postcode{76131},
  %\state{Baden-W{\"u}rttemberg},
  \country{Germany}}}

\address[4]{\orgdiv{Institute of Optronics, System Technologies and Image
  Exploitation (IOSB)}, \orgname{Fraunhofer},
  \orgaddress{
  %\street{Fraunhoferstra{\ss}e 1}, 
  %\postcode{76131},
  %\state{Baden-W{\"u}rttemberg},
  \country{Germany}}}

\address[5]{\orgdiv{Center for Machine Learning (CML)}, \orgname{Fraunhofer},
  \orgaddress{
  %\street{Fraunhoferstra{\ss}e 1}, 
  %\postcode{76131},
  %\state{Baden-W{\"u}rttemberg},
  \country{Germany}}}

\address[6]{\orgdiv{Zoological Institute (ZOO)},
\orgname{Karlsruhe Institute of Technology},
\orgaddress{
%\street{Fritz-Haber-Weg 4}, 
%\postcode{76131},
\country{Germany}}}

\corresp[$\ast$]{Corresponding author.
  \href{mailto:nils.friederich@kit.edu}{nils.friederich@kit.edu}}
\corresp[$\oplus$]{Supervised Equally}

\received{Date}{0}{Year}
\revised{Date}{0}{Year}
\accepted{Date}{0}{Year}

\abstract{
\underline{Summary:} Promptable segmentation models turn one 3D bounding box into an instance mask. Box placement therefore limits how fast microscopy volumes with thousands of objects are annotated. Each box must be checked in several synchronized viewers, but napari-bbox shares no state between viewers. An edit must update every viewer and slows down with more boxes. napari-TurboBox instead holds one shared copy of the boxes for all viewers. In a four-viewer layout, napari-TurboBox needs a few milliseconds per mouse move to update the three slice viewers up to $10^4$ boxes. The same update in napari-bbox leaves the 33\,ms budget beyond about $10^3$ boxes and takes 159\,ms at $10^4$. Boxes stay axis-aligned and export to COCO and YOLO as per-slice 2D boxes.\\
\underline{Availability and Implementation:} napari-TurboBox is a pure-Python plugin for napari 0.6.6 to 0.9 at \url{https://github.com/N-Friederich/napari-turbobox} under the Apache~2.0 licence. Version 0.1.0 is archived at \url{https://doi.org/10.5281/zenodo.\zenodoDOI}.\\
\underline{Contact:} \href{mailto:nils.friederich@kit.edu}{nils.friederich@kit.edu}}


\maketitle
\thispagestyle{arxivfirst}
\pagestyle{arxivrun}

\begin{figure*}[t]
\centering
\includegraphics[width=0.98\linewidth]{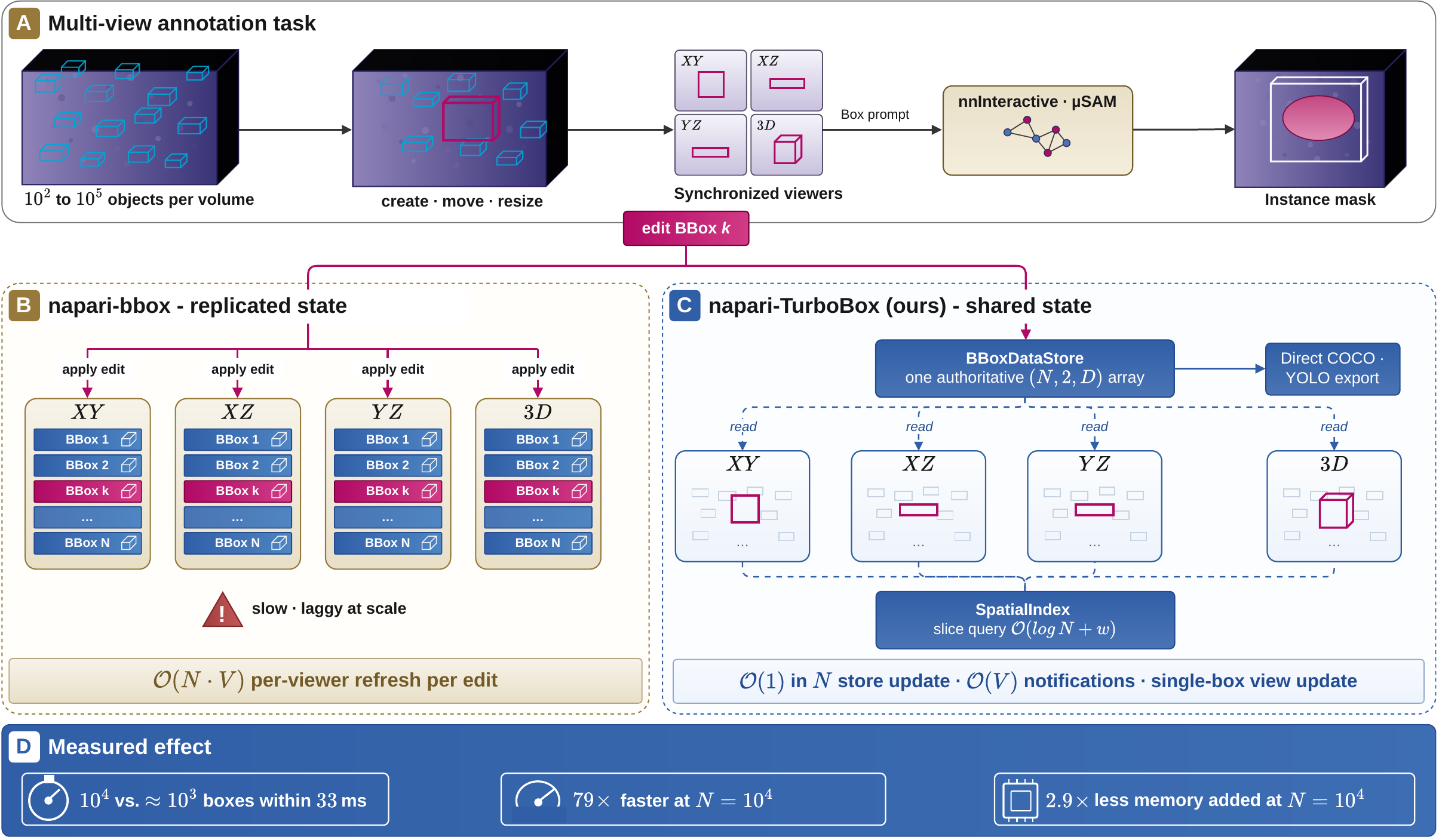}
\caption{\textbf{Faster multi-view bounding-box annotation with napari-TurboBox.}
\textbf{(A)} In multi-view 3D annotation, an axis-aligned bounding box (AABB) is created, moved or resized in one view, verified across the synchronized XY, XZ, YZ and 3D views, and used as a prompt for nnInteractive or $\mu$SAM to obtain an instance mask.
\textbf{(B)} napari-bbox keeps a separate layer with all $N$ boxes in each of $V$ viewers (replicated state); an edit must be applied to every layer at $\mathcal{O}(N)$ each, i.e.\ $\mathcal{O}(N \cdot V)$ per edit.
\textbf{(C)} napari-TurboBox writes the edited box once, in constant time, to a single authoritative $(N,2,D)$ store ($D{=}3$ spatial dimensions), notifies the $V$ views ($\mathcal{O}(V)$), and each view replaces only that box's shape; no per-viewer box array is copied. The shared state further supports constrained AABBs, slice-aware rendering and direct COCO/YOLO export.
\textbf{(D)} In the four-viewer benchmark, the range below the 33\,ms interaction threshold grows from about $10^3$ boxes with napari-bbox to $10^4$ with napari-TurboBox (about 2\,ms per update throughout). At $N=10^4$, napari-TurboBox is $79\times$ faster in median update time and its box-layer memory (single 2D viewer) is about $3\times$ lower.}
\label{fig:overview}
\end{figure*}

\section{Introduction}
\label{sec:intro}
Dense volumetric biomedical imaging routinely resolves hundreds to hundreds of thousands of objects per volume~\citep{amat2014,schmitz2017,nucmm,celegans_nuclei,celegans_atlas,weigert2020,cellseg3d}. Because 3D structures span multiple optical slices, manual voxel-level segmentation is prohibitively slow. Promptable deep learning models substantially reduce this burden: a single axis-aligned bounding box (AABB) enclosing a target structure provides sufficient spatial guidance to generate an instance mask~\citep{kirillov2023,ma2024} (Fig.~\ref{fig:overview}A). An imprecise box yields an imprecise mask~\citep{huang2024robox}, so accurate AABB placement directly determines downstream quality.

Placing a 3D box precisely is intrinsically a multi-view task. An object's $z$-extent is invisible in any single $XY$ slice yet decides whether the box encloses or clips the object, and depth in a projected 3D rendering is ambiguous under occlusion~\citep{tory2004}. Established volumetric annotation tools (e.g., ITK-SNAP~\citep{yushkevich2006}, webKnossos~\citep{boergens2017}, CVAT~\citep{cvat}) therefore commonly provide synchronized orthogonal slice views ($XY/XZ/YZ$) alongside a 3D rendering for live verification (Fig.~\ref{fig:overview}A). We call each such synchronized window a \emph{viewer} and use the four-viewer layout $XY$/$XZ$/$YZ$/3D throughout this work.

The napari viewer~\citep{napari}, widely used for Python-based bioimaging, can host several viewers side by side (synchronizing them is left to the application), with image data and annotations directly accessible as NumPy arrays~\citep{harris2020}.
This ecosystem increasingly integrates promptable deep learning models such as nnInteractive~\citep{isensee2025} and $\mu$SAM~\citep{archit2025} as interactive plugins. 
However, existing bounding-box tools like napari-bbox~\citep{napari_bbox} share no state across viewers, so each open viewer holds its own layer with the full coordinate state.
Consequently, translating or resizing a single box must be applied to the layer of every viewer, and each of these updates costs time that grows with $N$, i.e.\ $\mathcal{O}(N \cdot V)$ per edit. 
In a standard four-viewer layout ($V=4$), this overhead causes response times to exceed the interactive 33\,ms threshold ($\ge 30\,\text{Hz}$)~\citep{card1991,mackenzie1993,jota2013} beyond about $N = 10^3$ boxes, reaching about 160\,ms at $N=10^4$ (Fig.~\ref{fig:benchmarks}a).
To address these limitations, we present \textbf{napari-TurboBox},  a lightweight plugin built on napari's own \texttt{Shapes}  layer, designed for responsive multi-viewer 3D bounding box annotation via a shared-state architecture. A single in-memory store holds one authoritative copy of the boxes and propagates each edit to all viewers without duplicating the array. napari-TurboBox therefore needs about 2\,ms per mouse move to update the three slice viewers up to $10^4$ boxes, well within the 33\,ms budget. Dense volumes with thousands of objects can thus be edited as a whole rather than in crops, whose boundaries clip objects and remove the depth context a box must be checked against. Every bounding box remains a strictly valid AABB during interaction and can be directly exported on demand to standard COCO~\citep{lin2014} and YOLO~\citep{redmon2016} formats for downstream segmentation and detection workflows (Fig.~\ref{fig:overview}).

% ─────────────────────────────────────────────────────────────────────────────
\section{Implementation}
\label{sec:implementation}

\begin{figure}[t]
\centering
\includegraphics[width=0.48\textwidth]{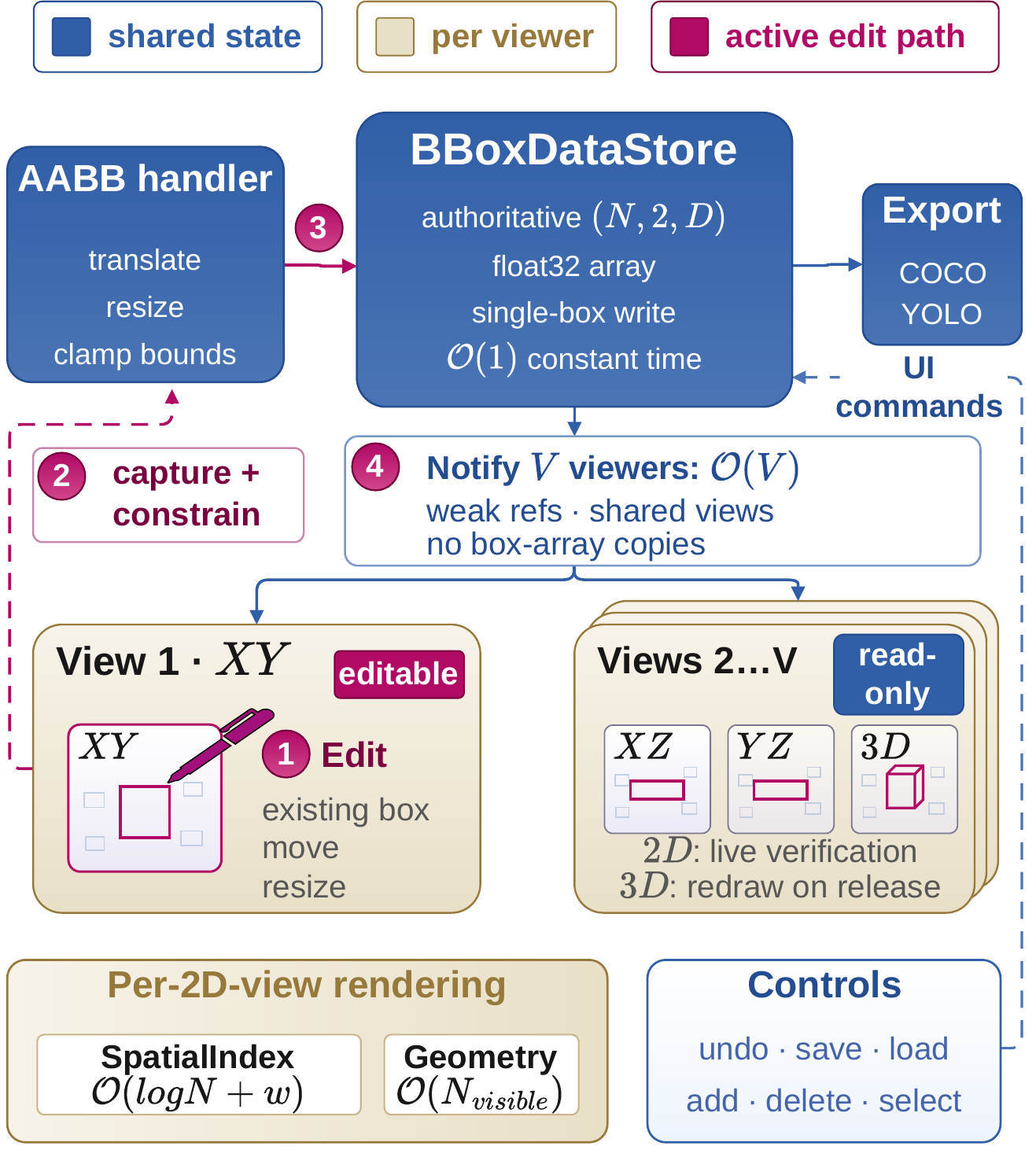}
\caption{\textbf{Shared-state architecture and numbered edit path of napari-TurboBox.}
\textbf{(1)} The user moves or resizes an existing AABB in the active XY view.
\textbf{(2)} The AABB handler captures the interaction, restricts it to translation or resizing, and clamps the coordinates to the image bounds.
\textbf{(3)} The validated box is written once to the authoritative $(N,2,D)$ \texttt{float32} array in constant time ($\mathcal{O}(1)$).
\textbf{(4)} The store notifies $V$ weak-referenced viewer listeners in $\mathcal{O}(V)$, allowing the views to consume shared state without copying a separate box array for each viewer. Each 2D view updates its own spatial index and replaces only the edited box's shape. After a slice change it queries the boxes of the new slice in $\mathcal{O}(\log N + w)$ and rebuilds their geometry. The 3D view redraws on release. COCO and YOLO annotations are exported directly from the authoritative store, while napari's Add Rectangle, Delete and selection tools act on the store, and the plugin adds undo/redo and box import/export.}
\label{fig:architecture}
\end{figure}

\textbf{Shared-state synchronization.} 
The core difference in napari-TurboBox is state centralization: instead of each viewer keeping an isolated copy of all annotations, every viewer references a single shared \texttt{BBoxDataStore} (Fig.~\ref{fig:overview}C, Fig.~\ref{fig:architecture}). 
This store owns one authoritative $(N, 2, D)$ \texttt{float32} array for all $N$ boxes across $D$ spatial dimensions. 
When a user moves or resizes a box, the store overwrites that box's corners in place. A 2D view changes at most four coordinates, so the write takes constant time ($\mathcal{O}(1)$) regardless of $N$. The store then notifies registered viewers via weak references, handing each a read-only view of the array instead of a copy.

Importantly, centralizing the data store decouples coordinate propagation from slice-specific geometry generation. 
In napari-bbox, every edit must be applied to the layer of each viewer, and each layer update costs time that grows with $N$, scaling as $\mathcal{O}(N \cdot V)$.
In napari-TurboBox, each view replaces only the shape of the edited box in place, so napari's re-derivation of the displayed shapes is skipped. When a box enters or leaves the slice of a view, the view adds or removes only the shape of this box. A view rebuilds the geometry of all visible boxes, in $\mathcal{O}(N_{\mathrm{visible}})$, only when its slice changes or such a single-box update is not possible. The time per update is therefore nearly independent of $N$ (about 2\,ms up to $10^4$ boxes, Fig.~\ref{fig:benchmarks}a) and grows with the number of views $V$.

To keep interaction responsive, viewers update according to their display type: 
Orthogonal 2D slice viewers ($XY$, $XZ$, $YZ$) update box intersections live on every mouse sub-event. 
In contrast, the 3D verification viewer, where reconstructing volumetric wireframe meshes is computationally expensive, defers redrawing until the mouse drag is released (Fig.~\ref{fig:architecture}). 
For programmatic tasks such as dataset loading or clearing annotations, a built-in context manager pauses listener notifications, executing a single bulk redraw upon completion.

\textbf{Vectorized geometry and spatial indexing.} 
Beyond fast data sharing, rendering speed depends on how quickly each 2D viewer can identify the boxes crossing its current slice plane. 
In napari-bbox, this requires checking every box in the entire 3D volume ($\mathcal{O}(N)$ scan) whenever the slice changes.
To avoid checking irrelevant boxes, napari-TurboBox maintains, per view, a \texttt{SpatialIndex} of sorted coordinate boundary arrays per axis. 
Using binary interval search, the index retrieves intersecting boxes and click targets in $\mathcal{O}(\log N + w)$ time, where $w \ge N_{\mathrm{visible}}$ counts the boxes within the largest box extent of the query, and updates incrementally during single-box edits rather than rebuilding from scratch. 
A property-based test checks that the incrementally updated index answers point and slice queries exactly like a full rebuild across random add, update and remove sequences.
A geometry engine then constructs the visible 2D rectangles, in one batched NumPy operation~\citep{harris2020} above 1{,}000 visible boxes and from a per-box cache below.
A numerical tolerance of $10^{-6}$\,voxels ensures that box faces aligning exactly with integer slice planes are rendered consistently without boundary artifacts.

\textbf{Constrained interaction and AI-ready export.} napari's built-in shape tools let the annotator drag individual vertices, which produces rotated, free-form polygons. Both the promptable models and the COCO and YOLO detection formats require axis-aligned boxes, so those shapes cannot be used downstream, and the annotator still has to specify a 3D box while only ever drawing in a 2D slice. napari-TurboBox therefore constrains every interaction so each edit stays a valid AABB, and keeps the stored boxes ready to read out directly as detection labels and as model prompts. A custom event handler intercepts each mouse drag before napari's default logic and maps the drag to a translation or resize only, so a rotated box can never arise. The annotator edits in one 2D viewer while the other viewers stay read-only, so verifying a box against the orthogonal slices cannot accidentally move that box. When the image shape is known, a 2D-to-3D extrusion lifts a rectangle drawn in one slice along the full extent of the hidden axis. The annotator then trims it in the same viewer after rolling its displayed axes to include the depth axis. The known image shape also limits editing: mouse edits are clamped to the image bounds, and boxes outside these bounds are rejected by \texttt{add\_boxes} and the \texttt{bounding\_boxes} setter, so boxes edited in the viewer never need repair before export. Export is a direct read-out of the store rather than a separate conversion step: boxes are formatted natively for standard object detection pipelines (COCO~\citep{lin2014}, YOLO~\citep{redmon2016}) by automatically reordering napari's $(z, y, x)$ coordinates to standard $(x, y)$ formats, writing each 3D box as one 2D box per slice it covers.
Consequently, the exported boxes can immediately serve as training labels for 2D object detectors or as per-slice box prompts for segmentation models (e.g., $\mu$SAM, nnInteractive).

% ─────────────────────────────────────────────────────────────────────────────
\section{Results}
\label{sec:results}

\begin{figure*}[t]
\centering
\includegraphics[width=0.98\textwidth]{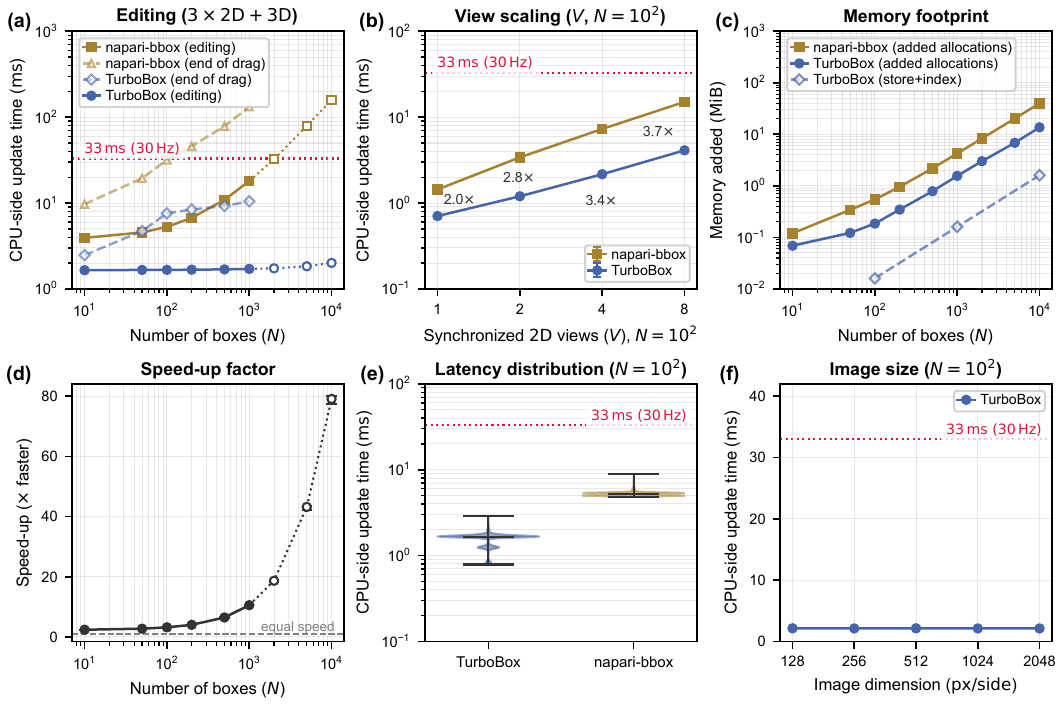}
\caption{\textbf{Benchmark of napari-TurboBox against napari-bbox.} Update times measured on the CPU in hidden napari viewers, both tools in one environment (one update = one mouse move propagated to all live 2D views). Panels a, b, d and e show $k{=}5$ runs per tool with 50 drags of 60 mouse moves on a 30\,Hz schedule per run. All drag measurements start with one untimed drag per configuration. Reported times include synchronization and each view's geometry update, no frame is rendered, and the 3D view and the thumbnails update once at the end of each drag for both tools (dashed, shown up to $N{=}10^3$). Open markers in (a) and (d): $N \ge 2{,}000$ from the large-$N$ run. Dotted red line: 33\,ms budget. \textbf{(a)}~Update time versus box count $N$ in the four-viewer layout (XY editable, XZ/YZ read-only, 3D on release). \textbf{(b)}~Update time versus number of synchronized 2D views $V$ at $N{=}10^2$, no 3D view (labels: speed-up). \textbf{(c)}~Memory added by loading the boxes into one hidden 2D viewer versus $N$, measured once per condition with Python's tracemalloc, and the napari-TurboBox store plus spatial index (dashed). \textbf{(d)}~Speed-up over napari-bbox versus $N$ with bootstrap 95\% confidence intervals. \textbf{(e)}~Distribution of all update times at $N{=}10^2$. \textbf{(f)}~napari-TurboBox update time versus background image size at $N{=}10^2$ (one run, four 2D views, 12 drags per size).}
\label{fig:benchmarks}
\end{figure*}

We benchmarked our tool to assess how the update time grows with the box count $N$ and whether it stays within the 33\,ms budget ($\ge 30$\,Hz) required for interactive editing (Fig.~\ref{fig:benchmarks}a). We measured in real napari viewers with hidden windows, so no frame is rendered, with both tools in one environment (napari 0.7.0, napari-bbox 0.1.1) on one Apple M2 Pro with 16\,GB RAM. Drawing a frame is left to napari and the GPU and depends on the display and its refresh rate, so hidden windows keep the measurement on the plugin code. An automated protocol replays the same seeded mouse drags in the editable $XY$ view of both tools through napari's mouse dispatch into each plugin's own drag handler. napari-bbox has no multi-viewer synchronization, so its other views follow every move through its public per-box methods (select, remove, add), with napari's layer data events deferred to the end of the drag, as napari-bbox's own drag does. Each drag translates or resizes a box in 60 moves on a 30\,Hz schedule, with the $XZ$ and $YZ$ views placed through the box, and each configuration starts with one untimed drag. Waiting time between moves is excluded and spent busy-waiting. In a control with idle waiting (three runs of ten drags per tool, reported in the repository), napari-TurboBox needed about 8\,ms per update in the four-viewer layout, and the speed-up grew from $1.1\times$ at $N{=}10$ to $2.3\times$ at $N{=}10^3$ and $19\times$ at $N{=}10^4$. Idle waiting did not change which of these configurations stayed within 33\,ms. We record the time from a mouse move until every live 2D view has updated and Qt has processed the pending events. Thumbnails and the 3D view are updated once, at the release, and reported separately. We sweep $N$ on blank $60{\times}256{\times}256$-voxel volumes across 50 drags and five replicate runs (separate processes) per tool, since the update cost tracks box and viewer count rather than image content (Fig.~\ref{fig:benchmarks}f). Moves within one drag are correlated, so the independent unit is the run. We take the median update time within each run and compare the tools on the five per-run medians, with speed-ups as ratios of medians and percentile-bootstrap 95\% confidence intervals from 10{,}000 resamples of the run medians (Fig.~\ref{fig:benchmarks}d). We call a configuration interactive when its median, 95th and 99th percentile (each taken as the median over the five runs) all stay within 33\,ms.

\textbf{Editing latency in the realistic layout.} In the four-viewer layout ($XY$/$XZ$/$YZ$/3D), the shared store substantially extends the interactive range (Fig.~\ref{fig:benchmarks}a). napari-bbox stays within the 33\,ms budget up to $N=10^3$ (median 18.0\,ms, 99th percentile 20.8\,ms), exceeds it by its 95th and 99th percentiles at $N=2{,}000$ (median 32.4\,ms, 95th percentile 33.5\,ms, 99th percentile 35.1\,ms) and needs 158.7\,ms per update at $N=10^4$. napari-TurboBox needs 1.6--2.0\,ms per update from $N=10$ to $N=10^4$ and keeps median, 95th and 99th percentile within the budget up to $N=10^4$ (99th percentile 4.4\,ms).
Its speed-up over napari-bbox is not a fixed factor but grows with $N$, from $2.4\times$ at $N{=}10$ to $3.2\times$ at $N{=}10^2$, $10.5\times$ at $N{=}10^3$ and $79\times$ at $N{=}10^4$ (Fig.~\ref{fig:benchmarks}d), because each napari-bbox view pays a layer update that grows with $N$ on every move, while napari-TurboBox changes one shape per view. With $10^4$ boxes, napari-TurboBox updates all 2D views in 2.0\,ms, whereas napari-bbox requires 158.7\,ms, i.e.\ about 6.3 updates per second.

\textbf{Scaling with viewer count.} Both tools show a roughly twofold increase in update time for each doubling of synchronized 2D views $V$ (Fig.~\ref{fig:benchmarks}b). At $N=10^2$, scaling $V$ from 1 to 8 raises the update time from 0.7\,ms to 4.1\,ms for napari-TurboBox and from 1.4\,ms to 15.1\,ms for napari-bbox. napari-TurboBox stays $2.0$--$3.7\times$ faster across all configurations because each napari-bbox view additionally pays a layer update that grows with $N$. The remaining per-view cost, updating each view's shapes and processing its events, is shared by both tools.

\textbf{Memory, 3D commit, and resolution.} At $N{=}10^4$, napari-TurboBox's box layer adds about $3\times$ less traced memory than napari-bbox's (13.7 vs 40.0\,MiB in one 2D viewer, Fig.~\ref{fig:benchmarks}c), of which the store and spatial index account for 1.6\,MiB. The ratio holds across $N$ ($2.7$--$2.9\times$ from $N=50$). Both tools defer the 3D wireframe update to the end of a drag, but napari-TurboBox triggers it automatically on release where napari-bbox relies on the application to do so, and the napari-TurboBox end-of-drag update (3D view and thumbnails) is $13\times$ cheaper at $N{=}10^3$ (10.5 vs 132.1\,ms). A box edit never touches the image layer, so image size cannot enter the update time of either tool. We verified this for napari-TurboBox, whose update time holds at 2.1\,ms (four 2D views) up to $2048^2$\,px (Fig.~\ref{fig:benchmarks}f).

\textbf{Application to real acquisitions.} 
While synthetic benchmarks isolate synchronization overhead, we verified end-to-end usability on real acquisitions of cellular nuclei and transcriptional clusters from confocal, electron, micro-CT, light-sheet and instant-SIM microscopy~\citep{celegans_nuclei,celegans_atlas,weigert2020,nucmm,nucmm_data,cellseg3d,cellseg3d_data,Hajiabadi_h,hajiabadi_data}.

Across all five modalities, 3D boxes derived from the instance labels (for instant-SIM, from a threshold segmentation) passed the plugin's validation unchanged, stayed valid under a replayed constrained mouse drag in the four-viewer layout and exported directly to COCO and YOLO, with one reproducible notebook per modality in the repository.

% ─────────────────────────────────────────────────────────────────────────────
\section{Conclusion}
\label{sec:conclusion}
napari-TurboBox makes dense 3D bounding box annotation practical inside napari: an annotator can place and check boxes on volumes with thousands of objects with a plugin update of a few milliseconds per mouse move, whereas napari-bbox leaves the 33\,ms budget beyond about $10^3$ boxes. By sharing a single authoritative state across all viewers instead of an independent box state per viewer, napari-TurboBox removes the update bottleneck at scale while reducing the memory of the box layer about threefold in a single viewer. Because every box is kept axis-aligned and exports directly to COCO and YOLO, the annotations are ready to train detectors or to prompt segmentation models such as nnInteractive without a conversion step. This turns box placement from a bottleneck into a routine part of preparing volumetric training data, and the shared-state architecture could be extended to other prompt types such as points and to direct prompt-response round-trips with foundation models inside the viewer.

% ─────────────────────────────────────────────────────────────────────────────
\section{Competing interests}
No competing interest is declared.

\section{Author contributions statement}
Conceptualization: NF, PK, MS, AS, LH, RM; Methodology: NF, AS, LH, RM;
Software: NF; Investigation: NF, PK; Resources: NF; Writing -- Original
Draft: NF; Writing -- Review \& Editing: all authors; Supervision: AS, LH,
RM; Project administration: LH, RM; Funding Acquisition: LH, RM.

\section{Acknowledgments}
The Helmholtz Association funds this project under the ``Helmholtz Imaging
Platform'', the authors N.~Friederich, A.~J.~Yamachui~Sitcheu, L.~Schilpp, S.~Allgeier, L.~Hilbert and R.~Mikut under the program ``Natural, Artificial and Cognitive Information Processing
(NACIP)'', and N.~Friederich, A.~J.~Yamachui~Sitcheu and T.~Lautenschlager through the graduate school ``Helmholtz Information \& Data Science School for Health
(HIDSS4Health)''.

Large language model tools (Claude Opus by Anthropic and ChatGPT Astra by OpenAI) assisted in writing and reviewing code, in checking the reported numbers against the benchmark records and in revising text. The authors reviewed and revised all content and take full responsibility for this publication.

\section{Data availability}
The source code, tests, tutorials and benchmark pipeline of napari-TurboBox are available under the Apache~2.0 licence at \url{https://github.com/N-Friederich/napari-turbobox}. The package is installed with \texttt{pip install napari-turbobox} from PyPI. The version described here is archived at \url{https://doi.org/10.5281/zenodo.\zenodoDOI}~\citep{turbobox_zenodo}. This archive also contains the benchmark records of all runs with their provenance files and the aggregated results (\texttt{paper\_results/}). The microscopy volumes used in the tutorials are publicly available~\citep{celegans_nuclei,nucmm_data,cellseg3d_data,hajiabadi_data}.

% ─────────────────────────────────────────────────────────────────────────────

\end{document}